\documentstyle[aps,prl,multicol,epsfig]{revtex}

\newcommand {\be}{\begin{equation}}
\newcommand {\ee}{\end{equation}}
\newcommand{\bey}{\begin{eqnarray}}
\newcommand{\eey}{\end{eqnarray}}

\begin{document}
\thispagestyle{plain}
\title{Chaotic Scattering of Microwaves}
\author{Tsampikos Kottos$^{\dag}$, Uzy Smilansky$^{\dag}$, 
  Joaquim Fortuny$^{\ddag}$ and Giuseppe Nesti$^{\ddag}$ \\[5pt]
  $^{\dag}$ Department of  Physics of Complex Systems, \\
  The Weizmann Institute of Science, Rehovot 76100, Israel \\
  $^{\ddag}$ Space Applications Institute and Joint Research Institute \\
  I-21020 Ispra (VA), Italy}
\date{accepted to Radio Science}
\maketitle

\vspace*{1cm}
\begin{abstract}
  In this paper we analyze a recent experiment conducted in an
  anechoic chamber, where the scattering of microwaves from an array
  of metallic cylinders was measured.  This is a system which displays
  chaotic scattering in the short wave limit. The analysis of the
  experimental data is aimed at elucidating the effects of the
  underlying chaos.  We describe a robust numerical method which
  provides the scattering matrix for any number of non overlapping
  reflecting cylinders.  We use this method to calculate the
  scattering from one, two and three cylinders, and to compare the
  results in the numerical simulations with those obtained in the
  experiment. Both simulations and measurements validate the presented
  theory.
\end{abstract}

\section{Introduction}
\label{sec:intro}
Scattering of electromagnetic waves in the microwave region is
governed to a large extent by the underlying geometrical optics or ray
dynamics, which provides the dynamical skeleton over which the
residual interference and diffraction effects can be superimposed. In
many instances, the ray dynamics is very complex, and it displays
chaotic features which earns this phenomenon the name ``chaotic 
scattering''. What makes the dynamics {\it chaotic} is that the
scattering trajectories which come from infinity, are trapped for 
a long time within the interaction region before they escape to
infinity. In the interaction domain, the motion is exponentially
unstable, and therefore the outgoing trajectories hardly ``remembers''
the initial conditions. This results in very complex reaction patterns,
fractal distributions, self similarity and other attributes which are
typical of classical chaotic dynamics. As is well known, the complex
patterns in dynamical systems are due to the existence of a ``strange
attractor'' in phase space. In chaotic scattering, which is not
a dissipative process, the ``strange attractor'' is replaced by a
``strange repeller'' which consists of all the trajectories which are
trapped forever in the interaction domain. The complexity of the ray
(or trajectory) dynamics manifests itself also in the corresponding
wave phenomena (be it electromagnetic or acoustic waves or all
possible de-Broglie --quantum-- waves describing electrons, atoms
or nuclei).  The purpose of the present paper is to illustrate the
importance of ``chaotic scattering'' in the context of electromagnetic
wave physics.

The system which we studied in this work consists of three parallel
metallic cylinders which scatter electromagnetic waves in the
microwave region. This study has been performed by means of
simulations and, what is more important, through measurements in the
anechoic chamber of the European Microwave Signature Laboratory
(EMSL)~\cite{emsl}. This system illustrates in an elegant and simple
way the ``generic'' features of chaotic scattering. It is also a
paradigm in the theoretical studies of this subject (see e.g.,
\cite{Eckhardt} \cite{GR89},\cite{USLesHouches89}, \cite{wirzba} and
references cited therein).

In the experiment, we measured the scattered fields in the TM mode
(with the electric field parallel to the cylinders axis). The
experimental set-up was planned such that one could neglect the
coupling to the other polarization component. The antennas were in the
far-field region of the cylinders (at a distance of about 10~m) and
therefore the wavefront can be considered to be planar. The incidence
is normal to the cylinders axis. This reduces the problem to the
treatment of a two-dimensional (2-D) scalar field scattering of three
cylinders in a planar geometry.

This paper is organized as follows. In Sect.~\ref{sec:formulation}, we
introduce the formulation of the problem and describe a robust
numerical method which provides the scattering matrix for any number
of non overlapping metallic cylinders.  We use this method to
calculate the scattering from one, two and three cylinders.  Then, in
Sect.~\ref{sec:rays}, we review the most important features which
characterize the chaotic ray dynamics in the present system.  The
identification of the fingerprints of chaotic scattering of the rays
in the actual experiments, where the scattering of waves of finite
wavelengths are measured, is addressed in Sect.~\ref{sec:semiclass}.
For this purpose we make use of the Eikonel (semiclassical)
approximation and propose a few correlation functions which are shown
to depend explicitly on functions which characterize the chaotic
dynamics of rays. In Sect.~\ref{sec:experiment}, we turn to the
description of the measurement set-up and the analysis of the
experimental data.  We compare the results in the numerical
simulations with those obtained in the experiment. We show the actual
correlation functions and compare them to the predictions of the short
wave theory proposed. We end this work with a summary of our
conclusions and a few suggestions for further experiments.

\section{Formulation of the Problem}
\label{sec:formulation}

\begin{figure}
  \epsfig{figure=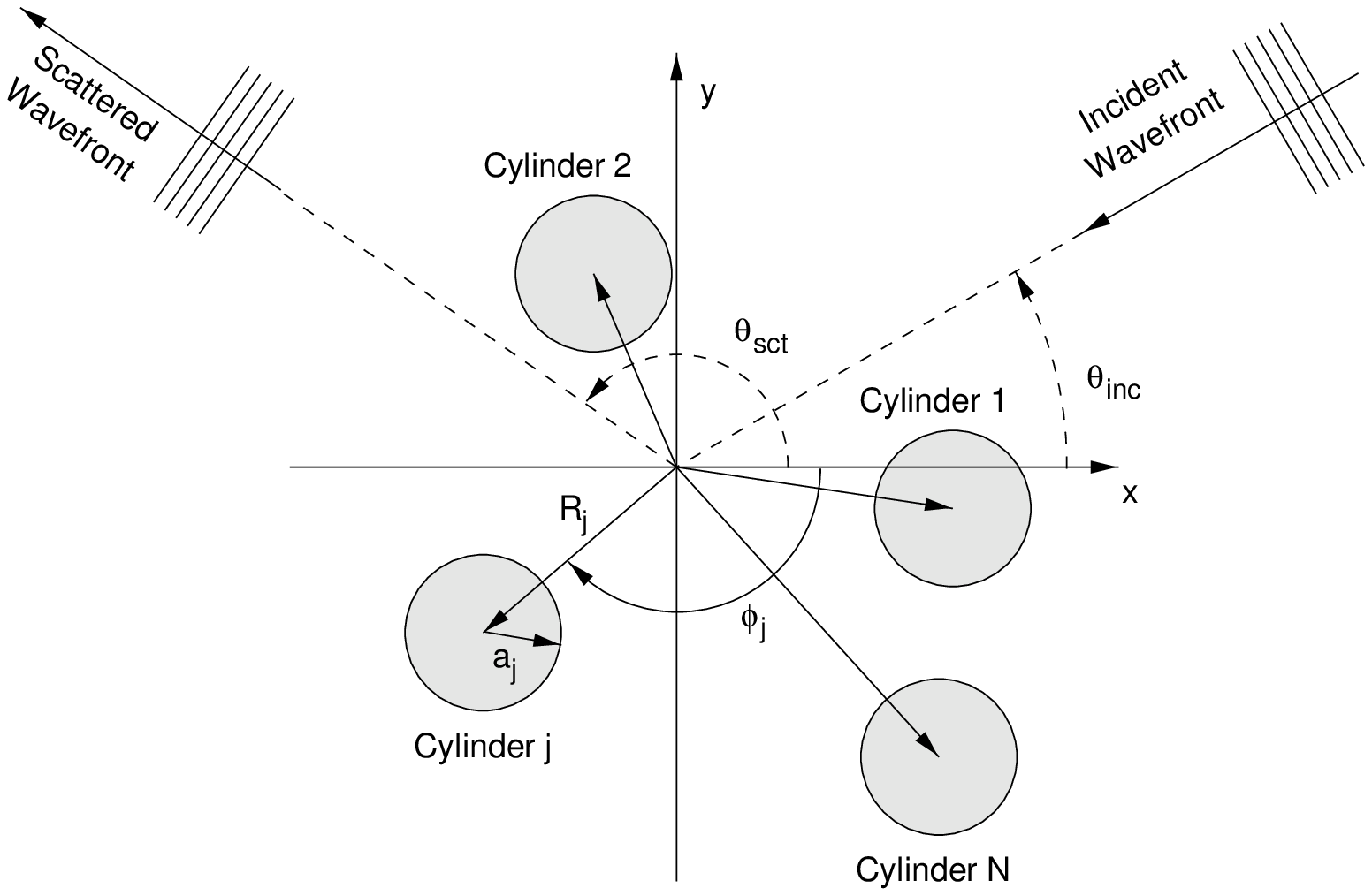,angle=0,width=18cm}
  \caption{Geometry of the problem.}
  \label{fig:geometry}
\end{figure}

Consider N parallel infinitely long conducting circular cylinders as
shown in Fig.~\ref{fig:geometry}. The axes of the cylinders are
parallel to the z-axis of a cylindrical coordinate system. The center
of the $j$th cylinder of radius $a_j$ is located at
($R_j,\phi_j$). A monochromatic plane wave with the electric field
parallel to the axes of the cylinders (TM mode) impinges at an angle
$\theta_{\rm{inc}}$. We assume a time dependence of the type
$\exp(-iwt)$.  The scattered field is scalar and must satisfy the
planar wave equation
\begin{equation}
\label{helm}
(\nabla ^2 + k^2)\Psi = 0
\end{equation}
where $k$ denotes the frequency wavenumber. This equation is solved
subject to Dirichlet boundary condition on the boundaries of the
cylinders (i.e. the total tangential electric field vanishes on the
surface of all cylinders). Thus
\begin{equation}
\label{bd}
\Psi(\bf{r})|_{\bf{r}=\bf{r_j}} = 0
\end{equation}
where $\bf{r_j}$ is a point on the boundary of the cylinder $j$. At
large observation distances, the solution may be decomposed into a
summation of incoming and outgoing cylindrical waves as follows,
\begin{eqnarray}
\label{asboun}
\Psi_{k,l}(\bf{r}) &\simeq& \frac{1}{(2\pi k r)^{1/2}} \times \\
& &\sum_{l'=-\infty}^{\infty}
\left [ e^{-i[kr-(l'\pi /2)-(\pi /4)]} \delta_{l,l'}
+  S_{l,l'} e^{i[kr-(l'\pi /2)-(\pi /4)]} \right ] e^{il'\phi}
\nonumber
\end{eqnarray}
which is used as a definition of the scattering matrix $S$.  This
matrix contains all the information needed for a complete description
of the scattering process and can be calculated exactly by making use
of the KKR method \cite{K47}. The various matrix elements $S_{l,l'}$
describe the scattering between states of different angular momenta
$l$ and $l'$. Because of flux conservation $S$ must be unitary.

Following the work of Gaspard and Rice \cite{GR89} we proceed to the
calculation of the $S$ matrix for the general case of $N$
non-overlapping cylinders. To this end, one uses the Green's theorem
in order to convert Eqs.~(\ref{helm}) and (\ref{bd}) into an integral
equation of the form
\begin{eqnarray}
\label{greent}
\int_D d^2r' \left[ \Psi_{k,l}(\bf{r}')(\nabla_{\bf{r}'}^2 +
  k^2)G_0({\bf r},{\bf r}') -
  G_0(\bf{r},\bf{r}')(\nabla_{\bf{r}'}^2 + k^2)
  \Psi_{k,l}(\bf{r}) \right] =  \\
\oint_{\partial D} dr' {\hat n(\bf{r}')} \left [ \Psi_{k,l}({\bf
    r}')\nabla_{\bf{r}'}
  G_0(\bf{r},\bf{r}') - G_0(\bf{r},\bf{r}') 
  \nabla_{\bf{r}'} \Psi_{k,l}({\bf r}') \right ].
\nonumber
\end{eqnarray}
Here, $D$ is the domain exterior to the cylinders but inside a circle
with a boundary $\partial D_{\infty}$ which is large enough to enclose
all the cylinders.  $\partial D$ is the boundary of this domain, and
the normal pointing outside is denoted by ${\hat n}(\bf{r}')$. The
boundary $\partial D$ of $D$ is the union of $\partial_{\infty} D$ and
the boundaries of the $N-$cylinders and $\partial_j
D,\,\,\,(j=1,\ldots,N)$.  The free (outgoing) Green function is given by
\begin{equation}
\label{gff}
G_0(\bf{r},\bf{r}') = -\frac{i}{4}H_0^{(1)}(k|\bf{r}-\bf{r}'|)
\end{equation}
where $H_0^{(1)}$ is the zeroth-order Hankel function of the first kind.

Choosing $\bf{r}$ on the $j$ cylinder the first part of
Eq.~(\ref{greent}) vanishes and we get:
\begin{equation}
  \label{qc1}
  0= \sum_{i=1}^{N} \oint_{\partial_j D} dr' 
  {\hat n(\bf{r_j}')} \left[ G({\bf r_j},\bf{r_j}')
    \nabla_{\bf{r}'}\Psi_{k,l}(\bf{r_j}')\right]
\end{equation}
where the contour integral has $N+1$ parts. By calculating this
integral \cite{GR89}, we obtain the gradient of the wavefunction at
the cylinder boundaries as
\begin{equation}
\label{matrix1}
A=CM^{-1}
\end{equation}
where the matrix $A$ defines the gradient of the wave function
\begin{equation}
  \label{Amatrix}
  {\hat n(\bf{r_j})} \nabla \Psi_{k,l}(\bf{r_j}) = 
  \sum_{m=-\infty}^{\infty}
  A_{ljm} e^{im\theta_j},\,\,\,\,l,m = 0,\pm 1,\pm 2,\pm 3,...
\end{equation}
and $\theta_j$ is the angular coordinate of the point $r_j$.

The matrix $M$ describes the multiple scattering between the $N$
cylinders and is found to be
\begin{equation}
  \label{Mmatrix}
  M_{jmj'm'} =  \delta_{mm'}\delta_{jj'} + (1-\delta_{j,j'})
  \frac{a_j}{a_{j'}}\frac{J_m(ka_j)}{H_{m'}^{(1)}(ka_{j'})} 
  H_{m-m'}^{(1)}(kR_{j,j'}) \;
  \zeta_{j,j'}(m,m')
\end{equation}
where $R_{j,j'}=R_{j',j}$ is the distance between the $j$ and $j'$
cylinder, and $a_j$ is the radius of the $j$th cylinder. Note that
$M$ has the structure of a KKR matrix and is the generalization of the
results of Gaspard and Rice \cite{GR89}. The matrix $\zeta$ is equal
to
\begin{eqnarray}
  \label{zeta}
  \zeta_{j,j'}(m,m')& = & \exp\left[i (m \, \phi_{j',j}-m'
    (\phi_{j,j'}-\pi))\right],\\
  \zeta_{j,j'}(m,m')& = & (-1)^{m-m'}  \zeta^*_{j',j}(m',m)
  \nonumber
\end{eqnarray}
and contains--besides a phase factor--the angle $\alpha_{j'j}$ of the
ray from the center of the cylinder $j$ to the center of the cylinder
$j'$ as measured in the local coordinate system of cylinder $j$. In
the simple case of one cylinder $M=\pi/2i$. Finally, the
matrix $C$ is given by
\begin{equation}
\label{cmatrix}
C_{ljm} = \frac{2i}{\pi a_j} e^{il\Phi_j} 
\frac{J_{l-m}(kR^j)}{H^{(1)}_m(ka_j)}.
\end{equation}
with $(R^j,\Phi_j)$ denoting the polar coordinates of the center of
the $j$th cylinder as measured in the global coordinate system.

If we now take $\bf{r}$ outside the smallest circle which encloses all
the cylinders, but inside the domain $D$, the first part of
Eq.~(\ref{greent}) provides $\Psi_{k,l}(\bf{r})$.  We then obtain
the $S$ matrix at large distance from the scatterer by propagating the
wave function, whose gradient on the cylinders is now known.  At the
last stage of the calculation we use the asymptotic expansions of the
Bessel functions at large distances $\bf{r}$ and the $S$ matrix is
identified by comparing with the asymptotic expression (\ref{asboun}).
The final result reads,
\begin{equation}
\label{smatrix}
S=I-iCM^{-1}D
\end{equation}
with the matrix $D$ defined:
\begin{equation}
\label{matrixd}
D_{ljm} = -\pi a_j J_{m-l}(kR^j)J_l(ka_j)e^{-im\Phi_j}.
\end{equation}

The method presented above has the advantage that the calculation of
the $S$ matrix is essentially reduced to the algebraic problem of
inverting the matrix $M$ defined in Eq.~(\ref{Mmatrix}). In
principle, this matrix is infinite. However, in order to carry on with
numerical calculations, we need to approximate it by a finite square
matrix of dimension $L$. The integer $L$ is chosen in such a way that
the truncated matrix is unitary to within the desired accuracy of the
calculation~\cite{GR89}.

The normalized scattered field is given in terms of the scattering
matrix $S$ as
\begin{equation}
\label{scampl}
f(\theta_{\rm{inc}},\theta_{\rm{sct}}; k) = 
\frac{e^{-i\pi/4}}{(2\pi k)^{1/2}}
\sum_{l,l'=-\infty}^{\infty}e^{-il(\theta_{\rm{inc}}-(\pi/2)}
(S_{l,l'}-\delta_{l,l'})e^{il'(\theta_{\rm{sct}}-(\pi/2)}.
\end{equation}
where $\theta_{\rm{inc}}$ and $\theta_{\rm{sct}}$ denote the
incidence and observation azimuth angles, respectively.

The 2-D bistatic radar cross section (RCS) is then simply computed as
\begin{equation}
\label{dcs}
\sigma_{\rm{bist}}(\theta_{\rm{inc}},\theta_{\rm{sct}}; k) = 
\left|f(\theta_{\rm{inc}},\theta_{\rm{sct}}; k)\right|^2.
\end{equation}

The presented formulation takes multiple scattering to any order and
is valid for any working frequency, provided that the number of
cylindrical modes used in the scattering matrix $L$ is sufficiently
large. In the next section, we will focus on the chaotic behavior of
the scattered fields in high frequency limit.

\section{Ray Dynamics}
\label{sec:rays}

Under the approximation of the geometrical optics, the electromagnetic
scattering is governed by the simple law of specular reflection at the
boundary of the cylinders. Given an arbitrary frame of reference, any
incoming ray can be identified by its direction $\theta_i$ and by the
nearest distance by which it approaches the origin - the impact
parameter $b_i$.  The impact parameter is proportional to the angular
momentum $l_i = k b_i$ where $k$ is the wavenumber. In the domain of
all possible initial rays $(\theta_i,b_{i})$, there is a compact set
of values for which the corresponding rays will impinge on one of the
cylinders. An initial ray in this relevant set will undergo multiple
reflections by the cylinders, until it is scattered away emerging as
the outgoing ray $(\theta_f,b_{f})$. The function
$\theta_f(b_i;\theta_i)$ where $\theta_i$ is kept fixed and $b_i$
varies within the relevant set is called the deflection function.

\begin{figure}
  \epsfig{figure=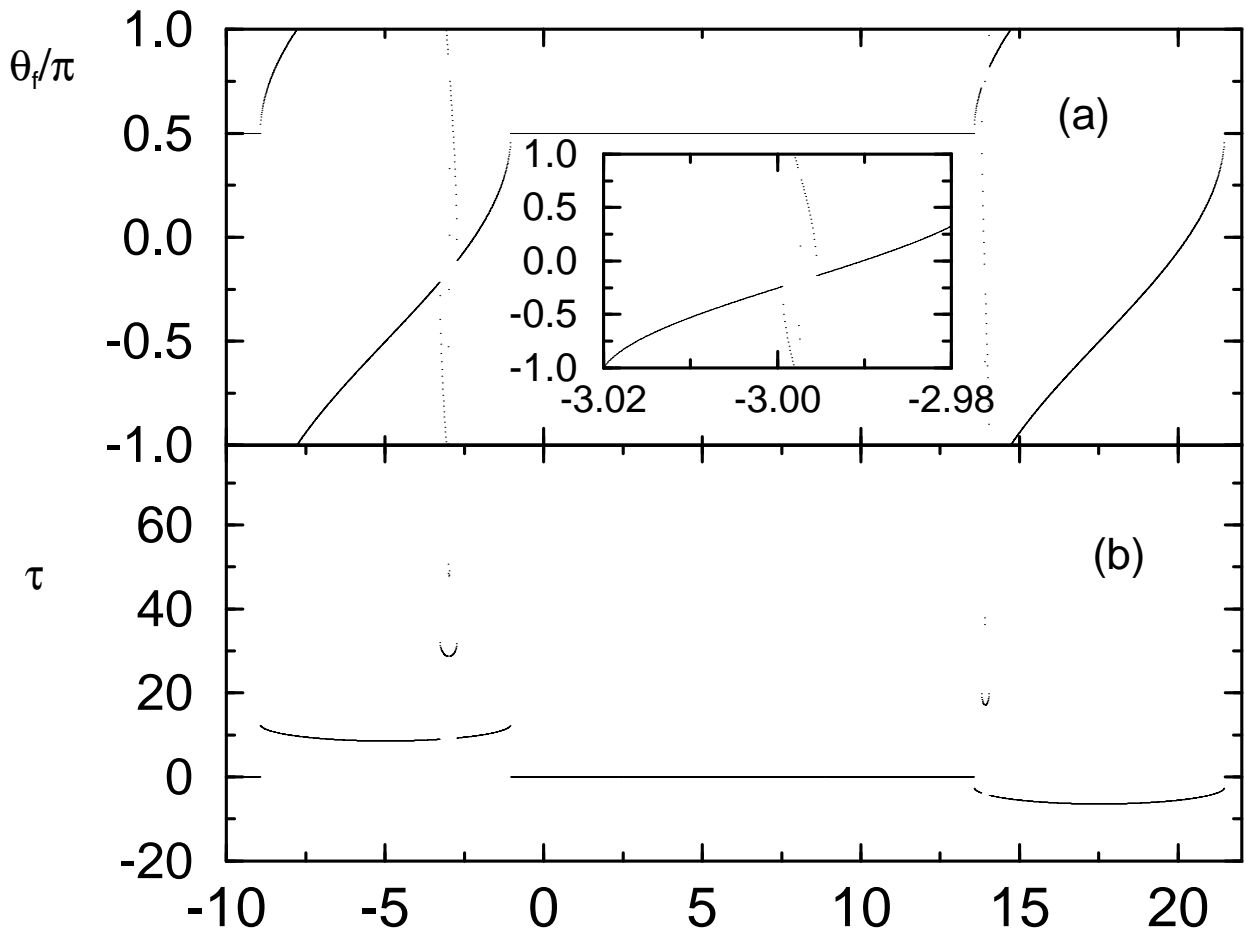,angle=-90,width=18cm}
  \caption{(a) The deflection function for the two cylinder system
  as a function of the impact parameter; (b) The corresponding dwell
  time as a function of the impact parameter}
  \label{fig1}
\end{figure}

For scattering on a single cylinder whose center is at the origin, we have
\begin {equation}
\theta_f(b_i;\theta_i)= \theta_i + 2\arcsin {b_i} \ \ \ \rm{for} \ \ \
|b_i| \le 1.
\label {single}
\end{equation}
The cylindrical symmetry implies $b_f=b_i$, and the actual deflection
$\theta_f - \theta_i$ is independent of $\theta_i$. Adding another
cylinder has two important consequences- the cylindrical symmetry is
broken, and multiple scattering becomes possible in the vicinity of
the trapped orbit which runs along the line which connects the two
centers of the cylinders.  Note that a ray which starts at infinity,
can approach the trapped ray arbitrarily closely, but it must finally
emerge. The resulting deflection function cannot be calculated
analytically.  Its complex structure is shown in Fig.~\ref{fig1}a.
The complexity is due to the fact that the trapped orbit is unstable.
(The instability emerges because the reflections are induced by
concave mirrors).  The longer the scattering orbits dwells next to it,
the more rapid are the fluctuations in the deflection function. The
concept of ``dwell time'' is essential for the following discussion and
it will be properly defined for a general scattering ray
(independently of the number of cylinders it may encounter). Consider
a scattering ray coming from $\theta_i,b_i$, and after $Q$ reflections
it goes out with $\theta_f,b_f$. Denote the reflection points on the
various cylinders by $\vec r_j$, with $1 \le j \le N$.  Then the dwell
time (measured in units of length) is defined as
\begin {equation}
 \tau(\theta_i,b_i) = {1\over c} \left \{\sum_{j=1}^{Q-1} |{\vec
r}_{j+1}-{\vec r}_j| +
{\vec r_1} \cdot {\vec n_i} - {\vec r_Q} \cdot {\vec n_f} \right \}.
\label{dwell}
\end{equation}
Here, $c$ is the speed of light, ${\vec n_i}$ and ${\vec n_f}$ are
unit vectors in the incoming and the outgoing directions.  The last
two terms above, remove the arbitrariness in the definition of the
point at which the dwell-time stop-watch starts (and stops) to tick.
Fig.~\ref{fig1}b shows the dwell time which correspond to the
deflection function for two cylinders. The correlation
of the complexity of the deflection with the dwell time is evident.

\begin{figure}
  \epsfig{figure=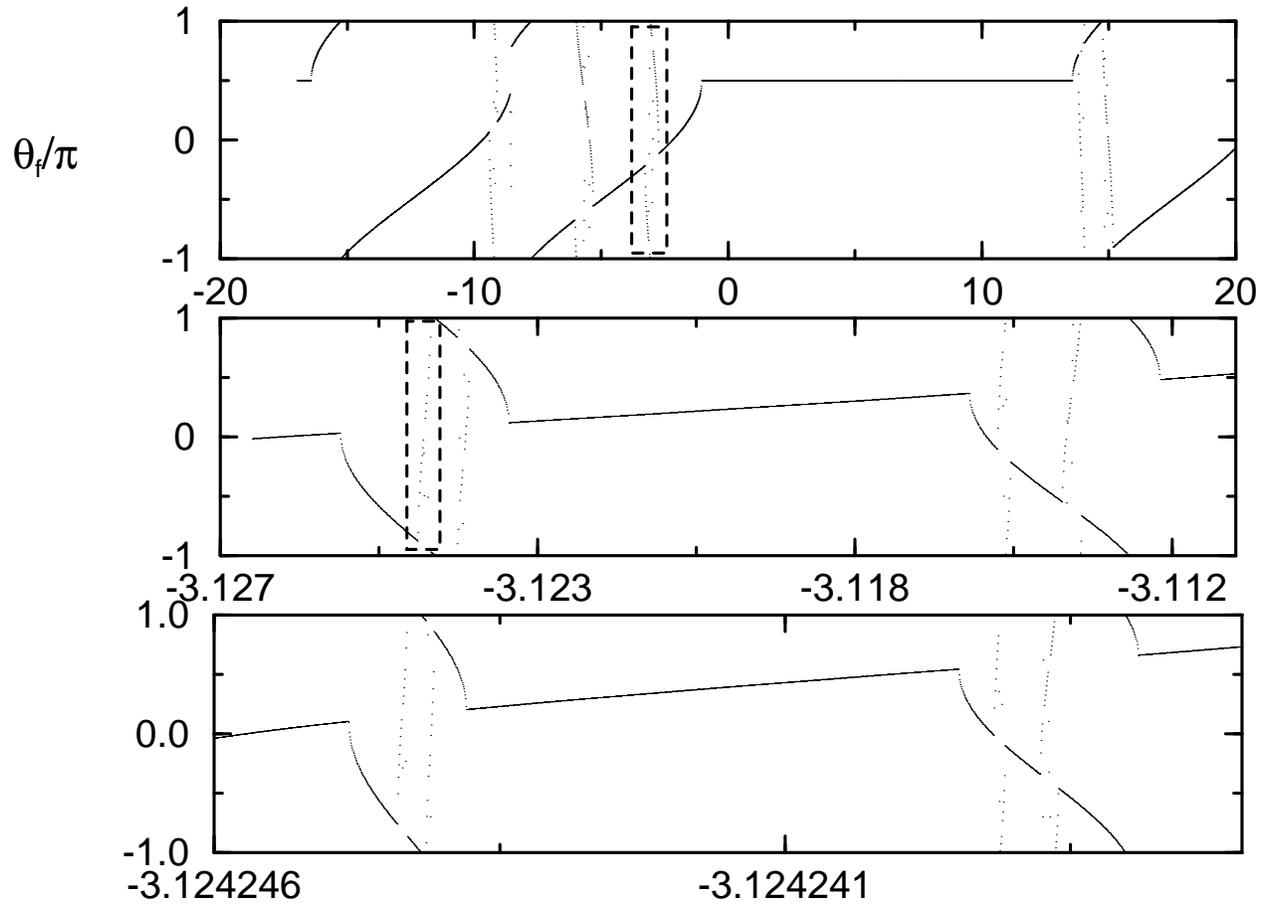,angle=-90,width=18cm}
  \caption{The deflection function for the three cylinder system
  as a function of the impact parameter. Three successive
  magnifications are shown, (the magnified domain is framed by dashes)
  and the fractal nature of the deflection function is obvious.}
  \label{fig2}
\end{figure}

\begin{figure}
  \epsfig{figure=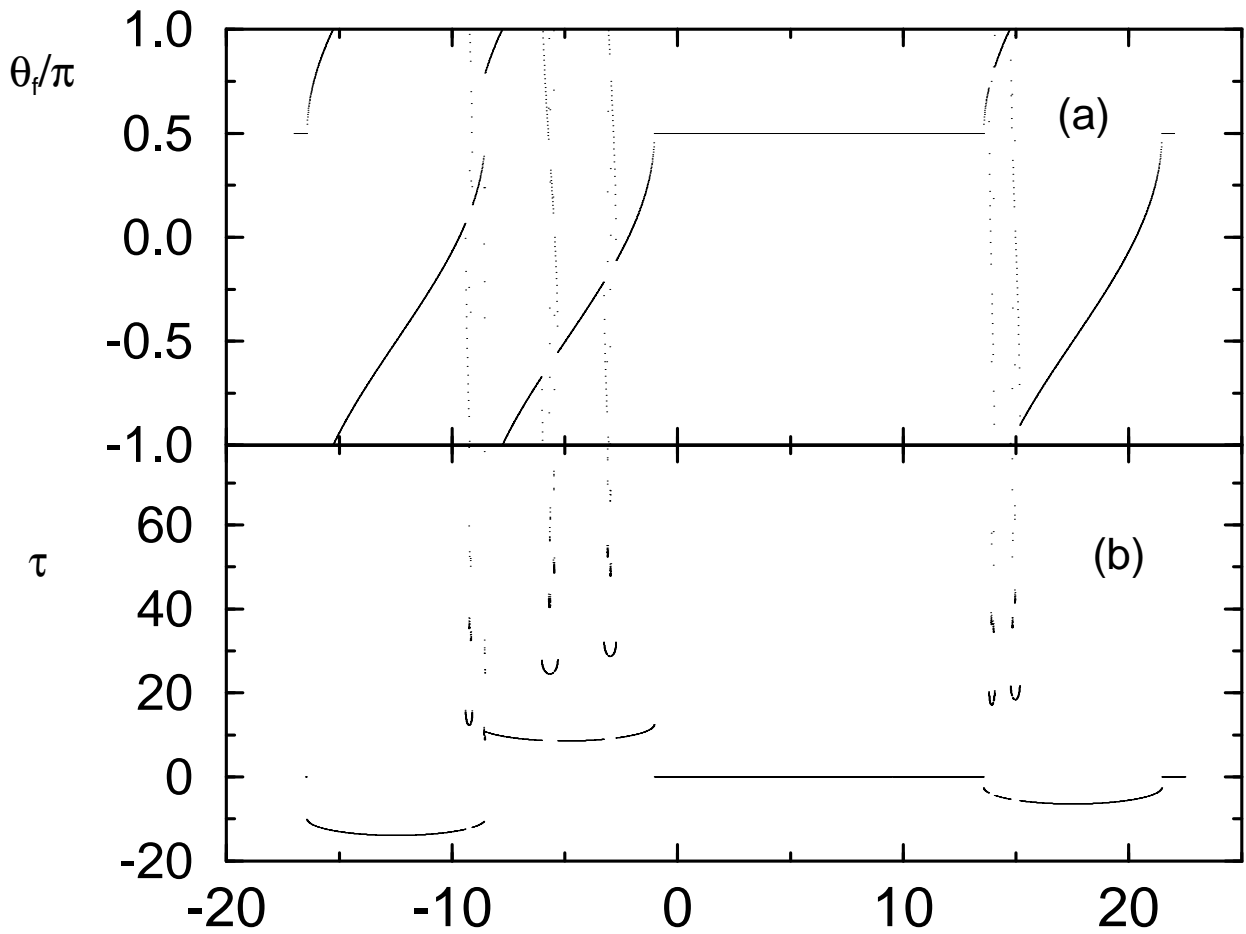,angle=-90,width=18cm}
  \caption{(a) The deflection function for the three cylinder for
  the entire range of impact parameter; (b) The corresponding dwell
  time exhibits singularities which are correlated with the
  self-similar structure observed in the deflection function.}
  \label{fig3}
\end{figure}

\begin{figure}
  \epsfig{figure=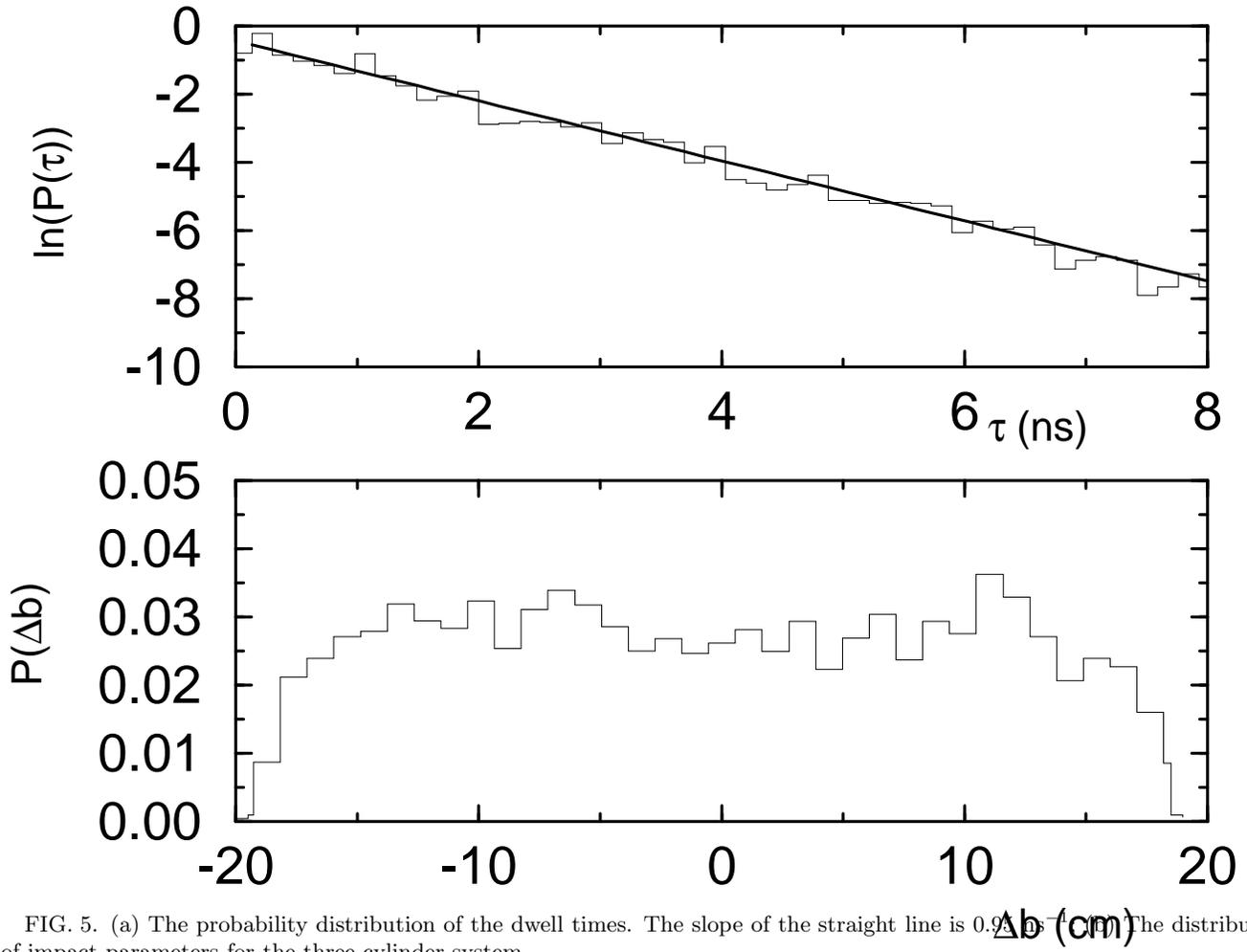,angle=-90,width=18cm}
  \caption{(a) The probability distribution of the dwell times. The
  slope of the straight line is $0.95~\rm{ns}^{-1}$; (b) The distribution
  of impact parameters for the three cylinder system.}
  \label{fig4}
\end{figure}

To generate chaotic scattering, one needs at least one more cylinder,
which should be placed such that no cylinder interrupts (shadows) the
lines which connect the boundaries of its neighbors. In this case,
there exists a Cantor set of trapped orbits which can be encoded by
the following symbolic dynamics: Denote the three cylinders by the
letters $A,B,C$. Then, to any infinite string of letters
$\{X_i\}_{i=1}^{\infty}$, $X_i \in \{A,B,C\} $, with $ X_i \ne
X_{i+1}$ there corresponds an unstable trapped orbit. If the string is
$N$ periodic $(X_i=X_{i+N}) $ for all $i$, the trapped unstable orbit
is periodic.  This is the ``strange repeller'' which is responsible
for the chaotic dynamics in our system: A ray which comes from
infinity and is scattered by the three cylinders is affected by an
infinite set of trapped orbits. It can dwell next to any of them
for some time, and then, due to the intrinsic instability, it can
either be trapped next to another member of the repeller, or scatter
out.  The resulting dynamics is much more complex than what we saw in
the two cylinders system, and it has the following features:
\begin{itemize}
\item The deflection function is singular on a Cantor set of $b_i$ values,
showing self
similar structures which occur on all scales (see Fig.~\ref{fig2}).
\item The dwell time function is singular on a Cantor set of $b_i$ values,
which is correlated with the singular, self similar structures observed in the
deflection function (see Fig.~\ref{fig3}).
\item The deflection function displays also smooth sections, which are due
to single scattering from the convex hull of the boundary. These trajectories
correspond to short dwell times, and we shall refer to this component of the
scattering system as ``direct''.
\item Excluding the direct component, the rest of the scattering dynamics is
ergodic - any  bundle of neighbor incident trajectories emerge with
 $(\theta_f,b_f)$
values which are uniformly distributed over the relevant set defined above.
\item The distribution of dwell times is asymptotically exponential (see
Fig.~\ref{fig4}a)
\begin {equation}
P_{\rm{dwell}}(\tau) \approx  \gamma_{\tau} \exp (- { \tau  \gamma_{\tau}})
\label {dwellP}
\end{equation}
and
\begin {equation}
\gamma_{\tau} =(1-d_H) \lambda
\end{equation}
where $d_H$ is the Hausdorff dimension of the Cantor set of singular
points on the $b_i$ axis, and $\lambda$ is the Lyapunov exponent which
characterizes the instability of the strange repeller.
\item The distribution of  impact parameter transfers (defined as the
difference between the final and the initial impact parameters) for the
ergodic component is approximately uniform
\begin {eqnarray}
P_{\rm{impact}}(b) \approx \left \{ \begin {array} {ll} {1\over D}
    & \mbox{if $|b| \le D/2$}
    \\ 0 &\mbox {otherwise} \end{array} \right.  \ \  .
\label{bP}
\end{eqnarray}
$D$ is an effective diameter of the part of the scatterer which induces
the ergodic scattering component (see Fig.~\ref{fig4}b).
\end {itemize}

Much more can be written on the mathematics and physics of chaotic ray
scattering.  The interested reader can find more details and
references in the articles cited above. The material presented above
suffices for the purpose of the following discussion.

\section{Correlation Functions -- The Fingerprints of Chaotic Dynamics}
\label{sec:semiclass}

The scattering amplitude $f(\theta_1,\theta_2;k)$ is the transition
amplitude to scatter from $\theta_1$ to $\theta_2$ at a wavenumber
$k$. In the Eikonel approximation the scattering amplitude can be
expressed as a superposition of amplitudes. Each amplitude corresponds
to a geometrical ray which is incoming from $\theta_1$ and is
out-going in the direction $\theta_2$. These rays are identified in
the following way. Consider the deflection function $\theta_f
(b_i;\theta_i)$ where $\theta_i$ is fixed at the value of interest
$\theta_1$ and one looks for the impact parameters $b_i^{s}$ which
satisfy the equation
\begin{equation}
\theta_2= \theta_f (b_i^s;\theta_1).
\label{boundarycond}
\end{equation}
The geometrical rays which are incoming at $(b_i^s,\theta_1)$ emerge
at $\theta_2$, and hence they correspond to the desired transition
$\theta_1\rightarrow \theta_2$. Due to the fact that the deflection
function oscillates wildly as a function of $b_i$, there is an
infinite set of values $b_i^s$ which satisfies (\ref {boundarycond}).
The Eikonel approximation for the scattering amplitude is
\begin{equation}
f(\theta_1,\theta_2;k) \approx  \left ( 2\pi k\right )^ {-{1\over 2}}
\sum_{s} \left [{\partial^2 \tau(\theta_1,\theta_2)
\over \partial \theta_1 \partial \theta_2}\right ]^{1\over 2} _s {\rm e}^{i
k \tau(\theta_1,
\theta_2)|_s + i\nu_s {\pi \over 2}}.
\label{sclamplitude}
\end{equation}
Here, the sum goes over all the rays which satisfy (\ref
{boundarycond}), and the dwell time for each ray (\ref{dwell}) appears
explicitly, and it is considered as a function of the parameters
$\theta_1,\theta_2$ which define the transition.  The pre-exponential
factors can be interpreted as the partial scattering amplitudes
brought by the individual rays. As a matter of fact, the sum of their
absolute squares is the cross section in the geometrical optics
approximation. It is obtained by deleting from the Eikonel cross
section all the terms which come from interferences of contributing
amplitudes.  The integers $\nu_s$ are the Maslov indices.

The scattering amplitude gets contributions not only from the multiply
reflected rays, but also from the ``direct" component of rays which
scatter from the convex hull of the cylinders. The statistical
treatment we propose is not applicable for this component, and we have
to subtract it from the measured or calculated data. This is done
using the following reasoning. The ``direct" component corresponds to
very short dwell times, and therefore its dependence on $k$ is very
smooth. Therefore, if one subtract the $k$ smoothed scattering
amplitude one remains with the desired statistical component. The
smoothing is done over a $k$ interval of the size $\Delta k$ centered
about $\bar k$. In the present case, the smoothing interval was taken
to be the entire interval where the measurement was carried out. Thus,
the ``statistical'' component of the scattering amplitude is
\begin{equation}
\tilde f(\theta_1,\theta_2;k) = f(\theta_1,\theta_2;k) - {1\over \Delta k}
\int_{\bar k -\Delta k/2}^{\bar k +\Delta k/2} f(\theta_1,\theta_2;k) {\rm
d}k \ \ .
\label{oscf}
\end{equation}

\begin{figure}
  \epsfig{figure=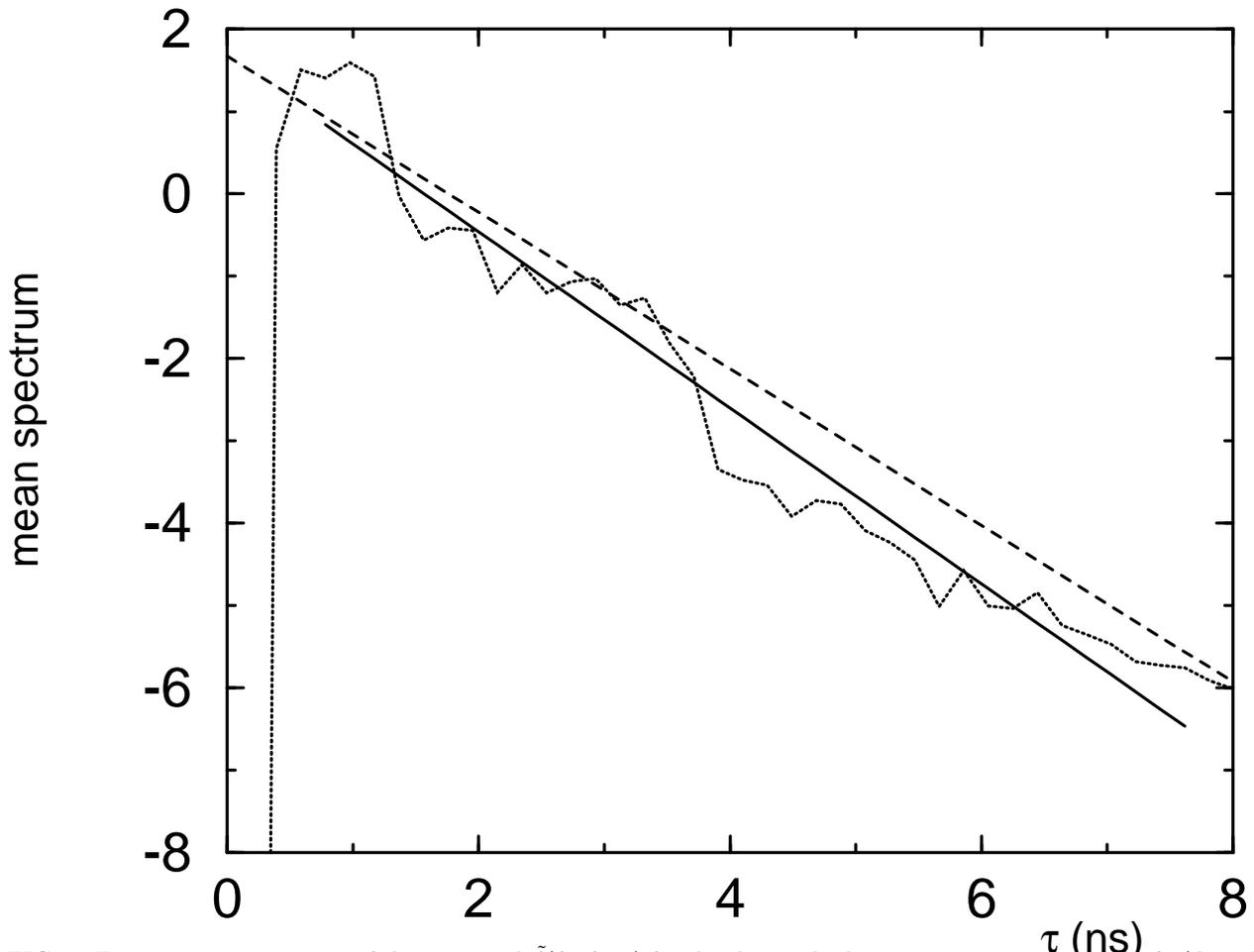,angle=-90,width=18cm}
  \caption{Fourier power spectrum of the measured $\tilde
  f(\theta_1,\theta_2;\nu)$ for the three cylinder system, averaged
  over $\theta_1, \theta_2$ (dotted line).  The best fit (bold line)
  and the the semi-classical prediction (dashed line) are shown.}
  \label{fig5}
\end{figure}

The $\tilde f(\theta_1,\theta_2;k)$ is the primary object of our
statistical study. Its Fourier transform with respect to $k$ provides
the spectrum of dwell-times for the rays which support the transition
$\theta_1 \rightarrow \theta_2$. When averaged over the incoming and
outgoing directions, it should coincide with the expression (\ref
{dwellP}) i.e.
\begin{equation}
\label{fft}
 \left < \left|\int dk
\tilde f(\theta_1,\theta_2;k) e^{-ik c \tau }\right|^2\right>_{\theta_1,
 \theta_2}
\simeq e^{-\gamma_{\tau } \tau }.
\end{equation}
A comparison between the experimental measurement and the classically
expected exponential decay is shown in Fig.~\ref{fig5}. The behavior is clearly
exponential over some 7 decades. The slope of the curve is equal to
$\gamma_{\tau}\approx 1.068\ ns^{-1}$, in reasonable agreement with
the value $\gamma_{\tau}=0.95\ ns^{-1}$, calculated from the ray
dynamics.

We will now define the angle averaged correlation function
\begin{equation}
C_{\theta}(\epsilon) =
\int_0^{2\pi}{{\rm d}\alpha \over 2\pi}
\left \langle \tilde f(\alpha ,\alpha+\Delta \theta ;k)
\tilde f^{*}(\alpha+\epsilon ,\alpha+\epsilon+\Delta \theta;k) \right
\rangle_k  \ .
\nonumber
\label{corr1}
\end{equation}
Here, we consider a situation where the outgoing angle relative to the
incoming direction $\Delta \theta$ is kept constant, and one
correlates the scattering amplitudes for different orientations of the
target. The $k$ averaging is the same as defined in (\ref {oscf}).
Using the Eikonel approximation (\ref {sclamplitude}) and the
relations
\begin{eqnarray}
{\partial \tau(\theta_1,\theta_1) \over \partial \theta_1}= -  b_1
\nonumber \\
{\partial \tau(\theta_1,\theta_2) \over \partial \theta_1}=   b_2
\label{canonical}
\end{eqnarray}
we can approximate the correlation function by
\begin{equation}
C_{\theta}(\epsilon) \approx \left \langle \int_0^{2\pi}{{\rm d}\alpha
\over 2\pi}\sum_{s}
 p_s(\alpha,\Delta \theta)
{\rm e} ^{-i k \epsilon (b^{(s)}_1-b^{(s)}_2) }
\right \rangle_k \ .
\end{equation}
To derive this expression, we assumed that $\epsilon$ is small so that
the phase differences can be approximated by the leading order in
their Taylor expansion. Off-diagonal contributions are discarded due
to the $k$ smoothing, and the pre-exponential factors are expressed in
terms of the geometrical-optics partial cross sections
\begin{equation}
p_s(\alpha,\Delta \theta;k) = 2\pi k  \left | {\partial^2
\tau(\theta_1,\theta_2)
\over \partial \theta_1 \partial \theta_2}\right |_s \ \ {\rm with} \ \
\theta_1 = \alpha \ ; \
\theta_2 = \alpha +\Delta \theta.
\end{equation}
The sum is extended over all rays $s$. We can now collect together all
the partial cross sections which are due to rays for which the impact
parameter transfer $ (b^{(s)}_1-b^{(s)}_2)$ equals $ b$
\begin {equation}
P_{impact}(b;k)=\sum_{s} \int_0^{2\pi}{{\rm d}\alpha \over 2\pi}\sum_{s}
 p_s(\alpha,\Delta \theta;k)
\delta( (b^{(s)}_1-b^{(s)}_2) - b )
\end{equation}
With this definition we get
\begin{equation}
C_{\theta}(\epsilon)  \approx \left \langle \int_{-\infty}^{\infty}{{\rm d}
b P_{impact}(b;k)
\rm e} ^{-i \epsilon  k b } \right \rangle _k \ .
\label{corrtheta}
\end{equation}
Since we subtracted away the ``direct" contribution, we can use the
ergodicity of the chaotic component and estimate $P(b)$ by the
expression (\ref {bP} ).

In a similar way we can define the cross correlation
\begin{equation}
X_{\theta}(\epsilon) =\\
\int_0^{2\pi}{{\rm d}\alpha \over 2\pi}
\left \langle \tilde f(\alpha ,\alpha+\Delta_1 \theta ;k)
\tilde f^{*}(\alpha+\epsilon ,\alpha+\epsilon+\Delta_2 \theta;k) \right
\rangle_k  \ .
\nonumber
\label{cross}
\end{equation}
where, in contrast with (\ref {corr1}) we correlate scattering
amplitudes measured at two different detector configurations. The rays
which support the scattering in the two configurations are completely
unrelated, and therefore we expect
\begin{equation}
X_{\theta}(\epsilon) = 0.
\label{crosscor}
\end{equation}

Another useful correlation function can be defined by
\begin{equation}
C_{k}(\nu) =\\
\int_0^{2\pi}{{\rm d}\alpha \over 2\pi}
\left \langle \tilde f(\alpha ,\alpha+\Delta \theta ;k +\frac{\nu}{2c})
\tilde f^{*}(\alpha,\alpha+\Delta \theta;k-\frac{\nu}{2c}) \right \rangle_k
\ .
\nonumber
\label{corr2}
\end{equation}

By repeating a similar argument as above, one can express $C_{k}(\xi)$
as a Fourier transform of the dwell time distribution
\begin{equation}
C_{k}(\nu) \approx \int_{-\infty}^{\infty}{{\rm d} \tau   P_{dwell}(\tau)
\rm e} ^{-i \nu  \tau }  \ .
\label{corrk}
\end{equation}
The dwell time distribution $P_{dwell}(\tau)$ was discussed above, and the
expression (\ref{dwellP}) implies that the $|C_{k}(\nu)|^2$ is a Lorentzian
with a width which is  proportional to $\gamma_{\tau}$.

The two correlation functions obtained above are typical for
scattering systems for which the resonances are overlapping and which
display therefore Erickson type fluctuations \cite{USLesHouches89}.
One of the most important achievements of the theory of chaotic wave
scattering was to recognize the fact that Erickson fluctuations are
the hallmark of the underlying chaotic ray dynamics.

Several other correlation functions (e.g., Hanbury-Brown Twiss type
correlations) can also be computed \cite{USLesHouches89}. However, the
measurement requires the use of two receiving antennas at close
(angular) proximity. This experiment was not yet carried out.

\section{Measurement Description and Data Analysis}
\label{sec:experiment}

Before comparing our numerical simulations with the experimental
measurements we shall give a short description of the experimental
setup. The measurements were performed in the anechoic chamber of the
European Microwave Signature Laboratory. This chamber has a
hemispherical shape with a diameter of about $20$~m.  It is equipped
with two separate TX/RX antenna modules, which can be moved
independently along a circular arch in a vertical plane, as well as
$27$~receiving antennas which are on the uniformly distributed
hemispherical dome.  The range from the center of the chamber to all
the antennas is the same: $9.56$~m. The system is fully polarimetric and
operates in the stepped frequency mode. 

\begin{figure}
  \epsfig{figure=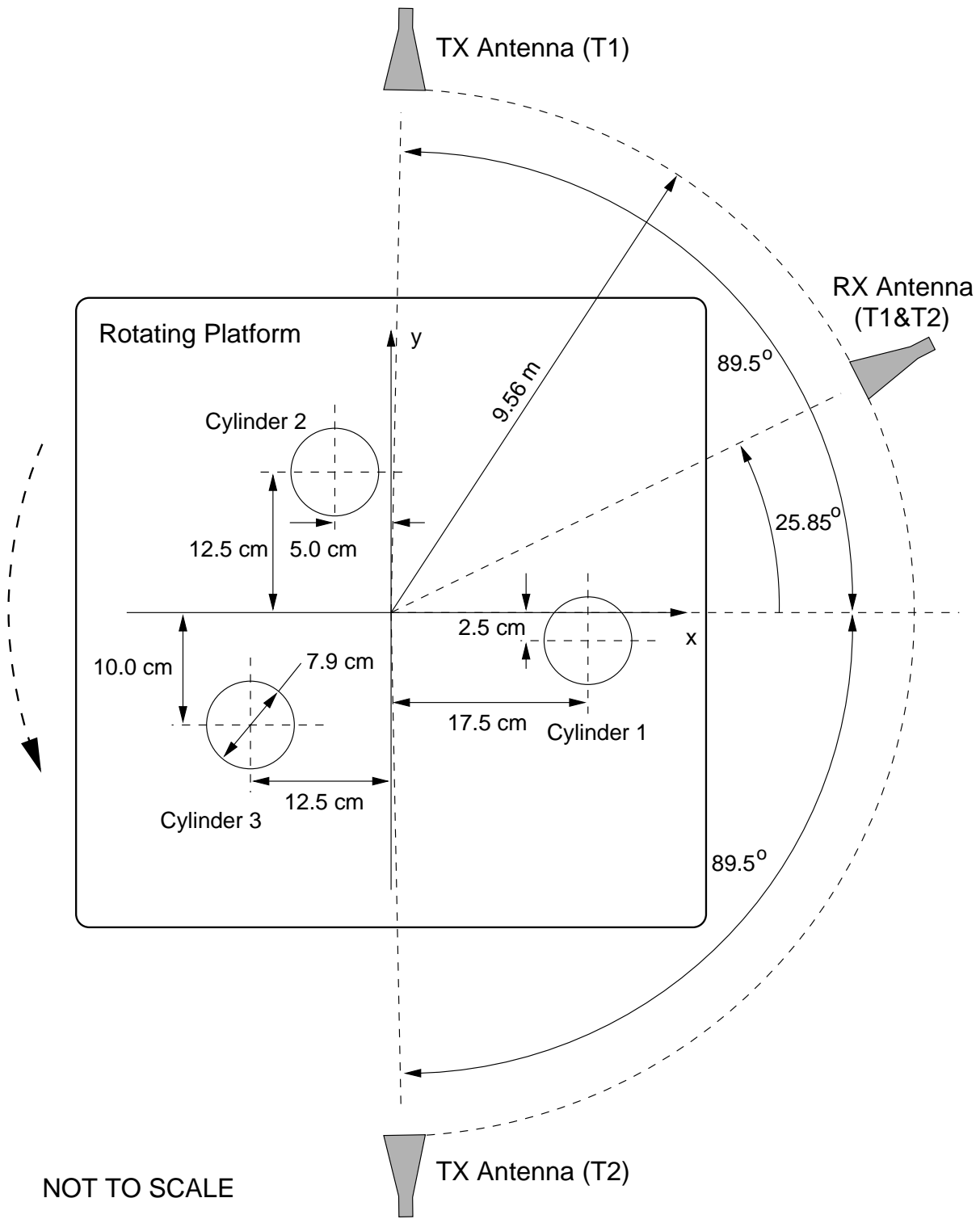,angle=0,width=16cm}
  \caption{Sketch of the measurement set-up used in the experimental 
    validation.}
  \label{fig:meas-geom}
\end{figure}

The measurement set-up used in the experimental validation is shown is
Fig.~\ref{fig:meas-geom}.  The target (consisting of one, two or three
identical copper cylinders of height $30$~cm and diameter $7.9$~cm)
was placed on a rotating table whose axis of rotation is parallel to
the axes of the cylinders. The center of gravity of the
three-cylinders coincides with the rotation axis (vertical axis).  Two
different antenna configurations were used, and they are referred to
as $T1$ and $T2$, with
\begin{itemize}
\item $T1$ configuration: TX at angle $89.5^o$; RX at angle $25.85^o$
\item $T2$ configuration: TX at angle $-89.5^o$; RX at angle $25.85^o$.
\end{itemize}

For each configuration, the bistatic scattered fields complex
amplitude was measured at $36$~equally spaced azimuth positions of the
table at $10^o$ intervals. The data were sampled within the frequency
interval $4-20$~GHz with a frequency step of $11.25$~MHz. The acquired
data in the frequency domain were empty room subtracted and gated in
the time domain in order to isolate the response of the cylinders from
the residual antennas coupling and eventual spurious reflections in
the chamber.  Then a bistatic calibration using a metallic sphere of
diameter $30$~cm placed at the focal point of the chamber, was applied
in the HH and VV polarizations.

\begin{figure}
  \epsfig{figure=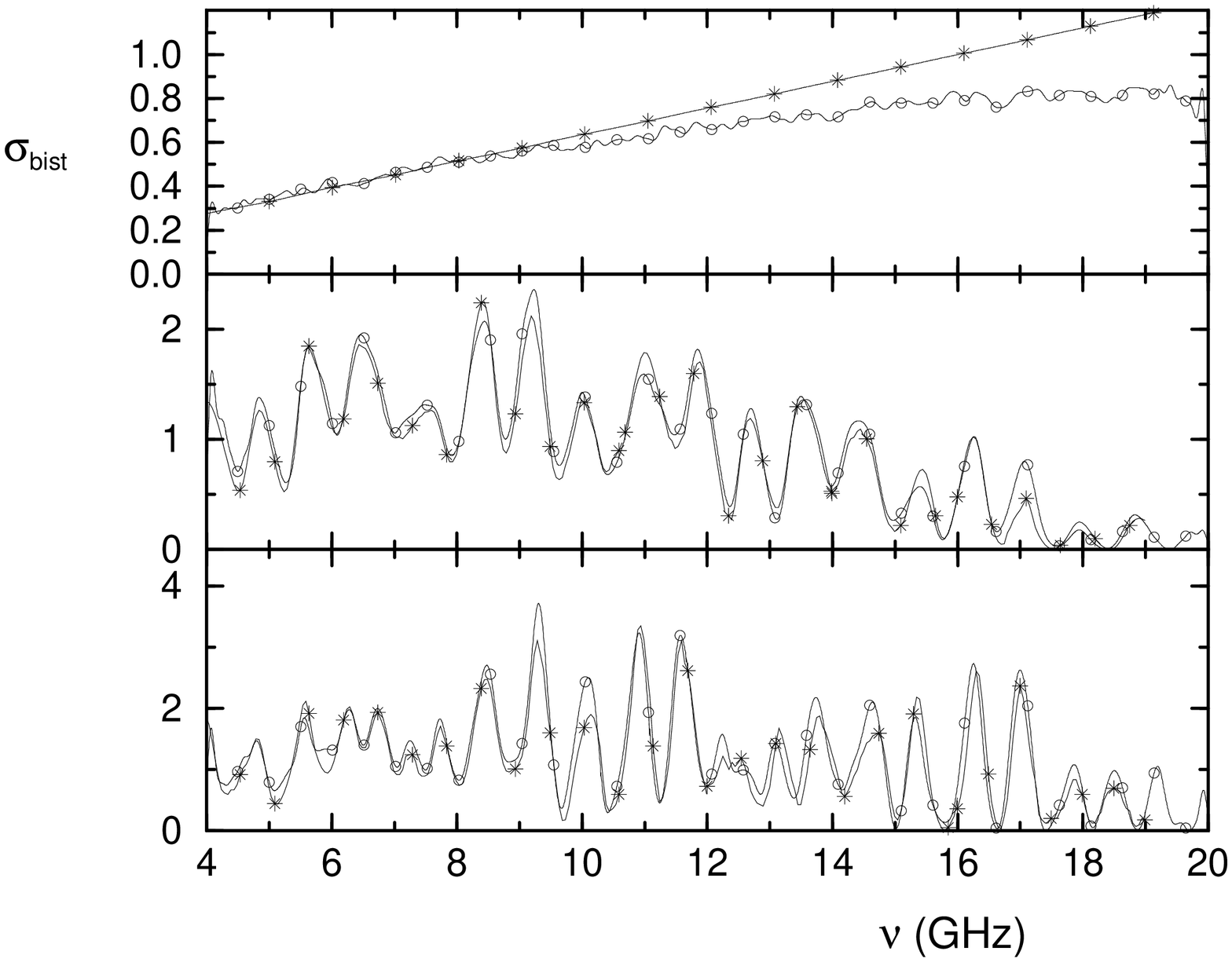,angle=-90,width=18cm}
  \caption{Experimental (circles) and numerical (stars)
  bistatic cross section vs.  frequency for the $T1$
  configuration.}
  \label{fig7}
\end{figure}

\begin{figure}
  \epsfig{figure=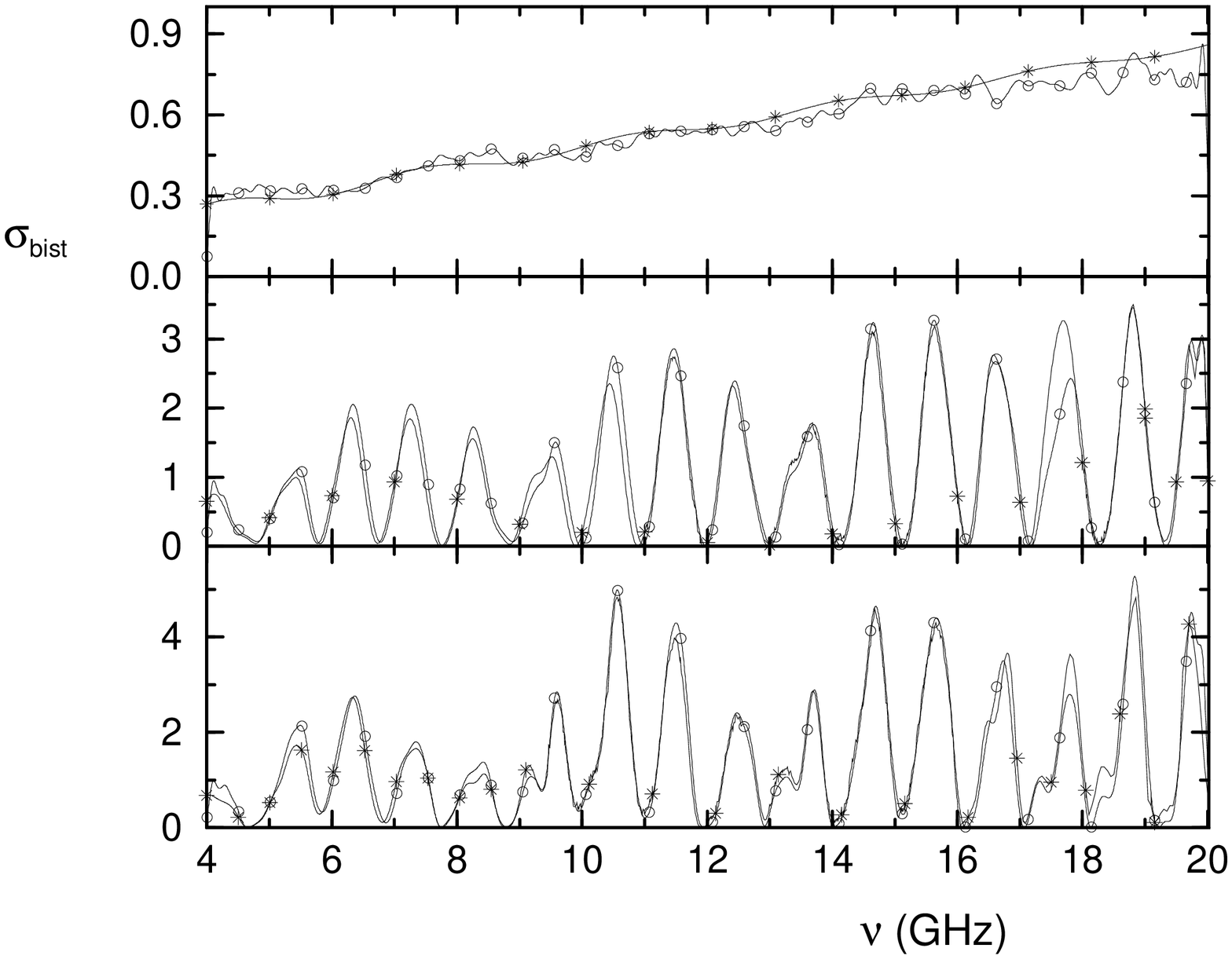,angle=-90,width=18cm}
  \caption{Experimental (circles) and numerical (stars)
  bistatic cross section vs.  frequency for the $T2$
  configuration.}
  \label{fig8}
\end{figure}

Some examples of experimental and numerical radar cross section vs.
the frequency $\nu$ are shown in Figs.~\ref{fig7} and \ref{fig8} for
antenna configurations $T1$ and $T2$ and for one, two and three
cylinders. As we see, the agreement of our numerical calculations with
the experimental measurements is quite good within the entire
frequency range.

We invested some effort to understand the deviations which appear for
the $T1$ configuration at the higher frequency range. In principle,
the scattering of a single cylinder should not depend on the table
orientation. However, the experimental data show distinct periodic
modulation with peak value of $\approx$ 15 \%. This can only be
explained as due to the finite geometry: the effective scattering
angle does depend on the orientation since the cylinder was not placed
at the center of the rotating table.

\begin{figure}
  \epsfig{figure=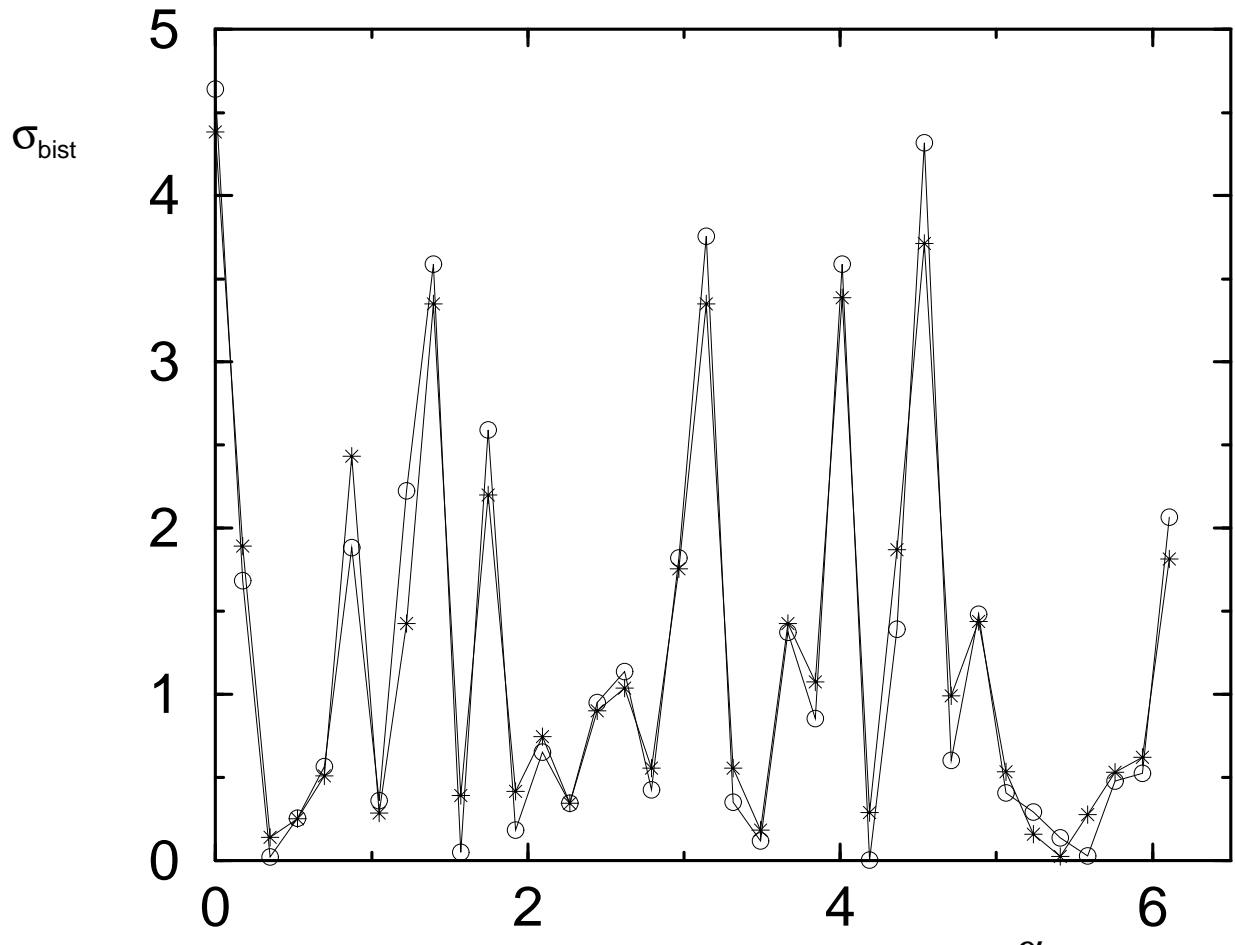,angle=-90,width=18cm}
  \caption{Experimental (circles) and numerical (stars)
    bistatic cross section vs. the azimuth position of the rotating
    platform $\alpha$ for the $T2$ configuration. The frequency is
    $10.485$~GHz.}
  \label{fig9}
\end{figure}

Finally, Fig.~\ref{fig9} shows the variations of the bistatic cross
section for some specific values of frequency, when the azimuth
position of the rotating table is varied. Again we are able to see a
quite good agreement with the experimental results which lead us to
the conclusion that our numerical calculations simulate quite well the
experimental data.

\begin{figure}
  \epsfig{figure=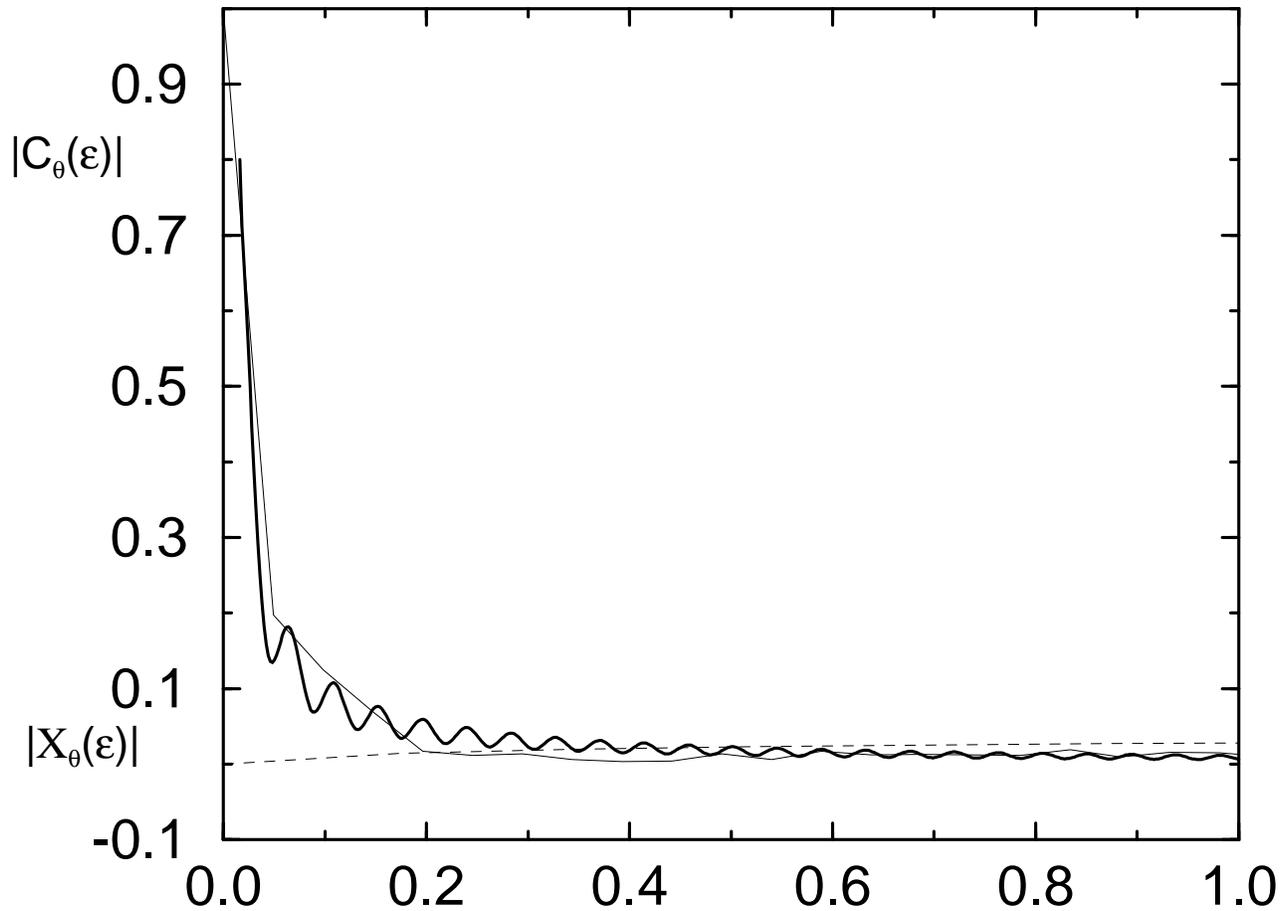,angle=-90,width=18cm}
  \caption{Numerical calculation of the correlation functions
    $C_{\theta}(\epsilon)$ (solid line) and $X_{\theta}(\epsilon)$
    (dashed line). The bold line correspond to the semiclassical
    prediction for $C_{\theta}(\epsilon)$. The antennas configuration
    is $T1$.}
  \label{fig10}
\end{figure}

We turn now to the statistical analysis of the scattering matrix. More
specifically we calculate the various correlation functions defined in
section \ref {sec:semiclass} in order to test the statistical
predictions. Fig.~\ref{fig10} displays the functions
$C_{\theta}(\epsilon)$ (\ref{corr1}) and $X_{\theta}(\epsilon)$ (\ref
{cross}).  The slight deviation of the cross correlation from the
expected value $0$ is due to the fact that we perform a finite
averaging over the frequencies.  The autocorrelation function, however
is consistent with Eq.~(\ref{corrtheta}).

\begin{figure}
  \epsfig{figure=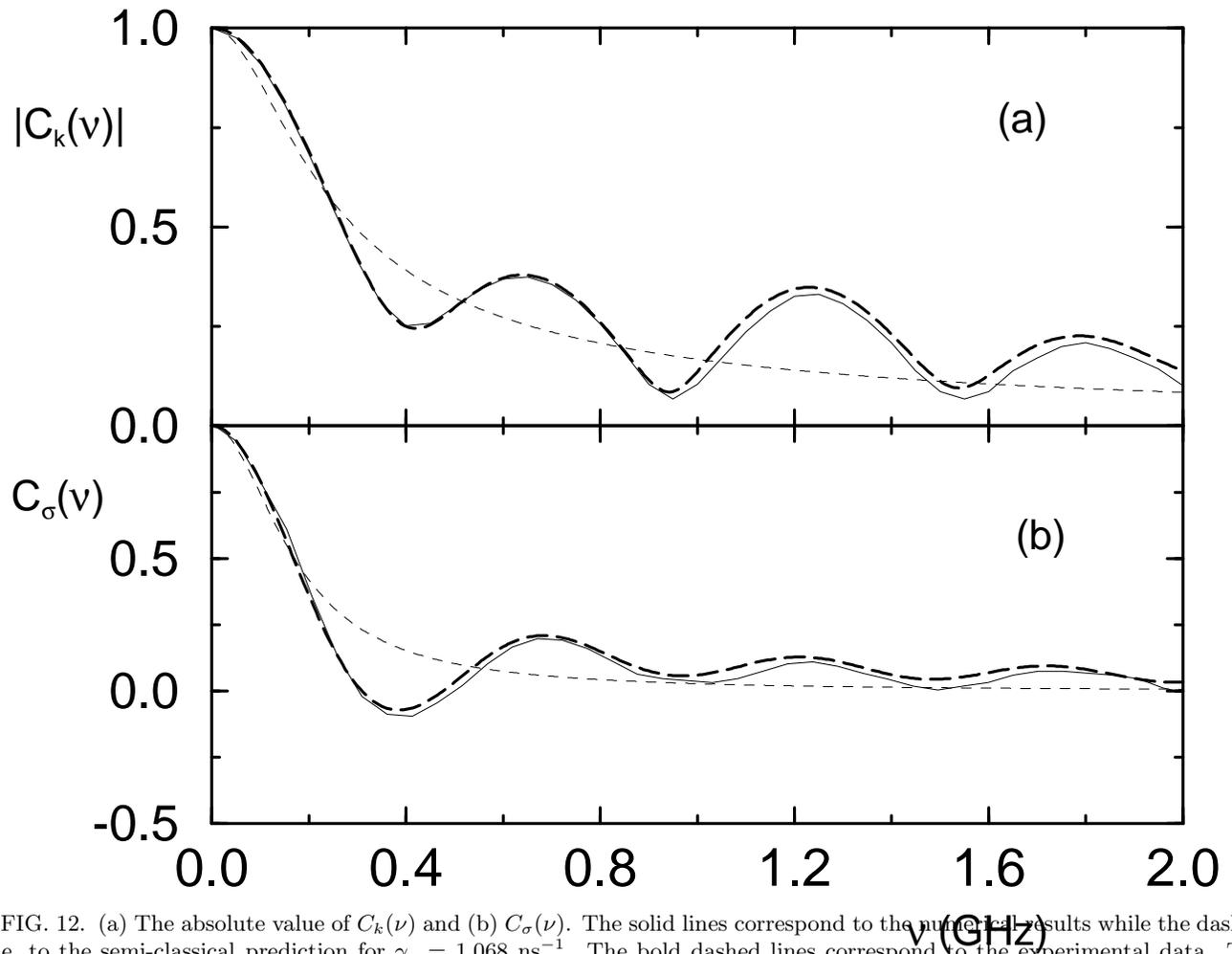,angle=-90,width=18cm}
  \caption{(a) The absolute value of $C_k(\nu)$ and (b)
    $C_{\sigma}(\nu)$.  The solid lines correspond to the numerical
    results while the dashed one, to the semi-classical prediction for
    $\gamma_{\tau}= 1.068~\rm{ns}^{-1}$. The bold dashed lines
    correspond to the experimental data. The antennas configuration is
    $T1$.}
  \label{fig11}
\end{figure}

Another interesting quantity we can evaluate is the energy correlation
function $C_k(\chi)$ defined in Eq.~(\ref{corr2}). From our
semiclassical consideration, we expect it to have a Lorentzian form
with half-width approximately equal to $\gamma_{\tau}$. Our results,
reported in Fig.~\ref{fig11}a are in agreement with the semiclassical
prediction.

A direct consequence of the above result, and of the underlying
ergodicity is that not only the amplitudes but also the cross sections
show similar correlations in energy. We expect
\begin{equation}
\label{ccros}
C_{\sigma}(\nu) \equiv \frac {\left \langle \hat {\sigma}_{\rm{bist}}
(k+\frac{\nu}{2c}) \hat
 {\sigma}_{\rm{bist}} (k-\frac{\nu}{2c}) \right \rangle _{k} }
{\left \langle \hat{\sigma}_{\rm{bist}}^2(k)\right \rangle_k} \simeq
\frac{1}{1+(\nu/\gamma_{\tau})^2}
\end{equation}
where $\hat {\sigma}_{\rm{bist}} (k) \equiv
\sigma_{\rm{bist}}(k) - \left \langle \sigma_{\rm{bist}}(k)
\right \rangle_k $ is the fluctuating part of the cross section.
Again, we calculate $C_{\sigma}(\nu)$ and recovered a satisfactory
agreement with the semiclassical predictions. The results are shown in
Fig.~\ref{fig11}b.

\section{Conclusions}

We have presented an algorithm that takes multiple scattering to any
order and we can use it to simulate any experimental configuration.
We present results which show that chaotic scattering can be
unambiguously identified in the experimental data.  Moreover, the
various theoretical simplifications which were assumed, as well as the
finite angular resolution in the experiment, did not blur the
signatures of the chaotic ray dynamics.

The main conclusion from the present calculations is that in the range
of frequencies used in the experiments, the effects due to chaotic 
scattering are prominent in the present system and, in general in most
systems in which electromagnetic radiation is scattered from metallic
objects. The scattered radiation field is so complex, that it shows
many features which would be typically attributed to a {\it random}
field.

One may investigate the subject further by performing e.g., the
measurements of the Hanburry-Brown Twiss effect, or by considering
other scattering systems such as cavities (see \cite{DoronUS}) or
waveguide networks (see \cite{kottos}). We also believe that the 
observations made here can be used, and should be taken into account
when heterodyn electromagnetic radiation experiments analyzed.

\newpage
\listoffigures
\newpage

\newpage

\end{document}